\documentclass[a4paper,10pt,twocolumn]{article}
\usepackage[utf8]{inputenc}
\usepackage{multicol,graphicx}
\usepackage{mathptmx}
\usepackage[T1]{fontenc}
\usepackage{textcomp}
\usepackage{url}
\usepackage[backend=bibtex, style=numeric,doi=false,isbn=false,url=false,maxnames=6,citestyle=numeric-comp]{biblatex}

\fontfamily{ptm}\selectfont

\bibliography{experimenter2018.bib}

\usepackage[font=small]{caption}
\usepackage{amsmath}

\newcommand{\mysection}[1]{\vspace{0.4cm} \uppercase{#1} \vspace{0.4cm}}
\newcommand{\mysubsection}[1]{\hspace{10pt}\textit{#1:}}

\usepackage{color}
\definecolor{orange}{rgb} {0.9,0.5,0.0}
\definecolor{green}{rgb} {0,0.5,0.3}

\begin{document}

\setlength{\textfloatsep}{10pt plus 1.0pt minus 2.0pt}	
\setlength{\columnsep}{1cm}


\twocolumn[%
\begin{@twocolumnfalse}
\begin{center}
	{\fontsize{14}{18}\selectfont
 \textbf{\uppercase{Would Motor-Imagery based BCI user training \\benefit from more women experimenters?}}\\}
 \begin{large}
 \vspace{0.6cm}
 A. Roc\textsuperscript{1, 2}, L. Pillette\textsuperscript{1, 2}, B. N'Kaoua\textsuperscript{3}, F. Lotte\textsuperscript{1, 2}\\
 \vspace{0.6cm}
 \textsuperscript{1}Inria Bordeaux Sud-Ouest, Talence, France\\
 \textsuperscript{2}LaBRI (University of Bordeaux / CNRS / Bordeaux-INP), Talence, France\\
 \textsuperscript{3}Handicap, Activity, Cognition, Health, University of Bordeaux, Bordeaux, France\\
 \vspace{0.5cm}
 E-mail: aline.roc@inria.fr, lea.pillette@inria.fr
 \vspace{0.4cm}
 \end{large}
\end{center}	
\end{@twocolumnfalse}%
]%





ABSTRACT: Mental Imagery based Brain-Computer Interfaces (MI-BCI) are a mean to control digital technologies by performing MI tasks alone. Throughout MI-BCI use, human supervision (e.g., experimenter or caregiver) plays a central role. While providing emotional and social feedback, people present BCIs to users and ensure smooth users' progress with BCI use. Though, very little is known about the influence experimenters might have on the results obtained. Such influence is to be expected as social and emotional feedback were shown to influence MI-BCI performances. Furthermore, literature from different fields showed an experimenter effect, and specifically of their gender, on experimental outcome. We assessed the impact of the interaction between experimenter and participant gender on MI-BCI performances and progress throughout a session. Our results revealed an interaction between participants gender, experimenter gender and progress over runs. It seems to suggest that women experimenters may positively influence participants' progress compared to men experimenters. 

\mysection{Introduction}

Brain Computer Interfaces (BCI) enable their users to interact with technologies by extrapolating their intentions from their brain activity, often measured using an electroencephalogram (EEG). In this article, we focus on Motor-Imagery based BCI (MI-BCI), for which users express their intentions by performing Motor Imagery (MI) tasks, such as imagining hands or feet movements, to induce change in their EEG signals and thereby control the BCI. MI-BCI represent promising new technologies. For example, they have proven effective for motor rehabilitation post-stroke and to interact with a variety of automated system, such as orthoses or video games \cite{Clerc16}.

Though, currently, MI-BCI do not enable a sufficient accuracy in detecting which task is performed by users \cite{Lotte13c}. Indeed, on average when trying to differentiate one MI task between two from EEG signals, a task is correctly recognized 75\% of the time \cite{Allison10}. Users need to train to acquire MI-BCI skills. Yet, around 10 to 30\% of users are unable to control MI-BCI \cite{Allison10}. Therefore, MI-BCI are still mainly at the development stage in research laboratories.

During MI-BCI experimental protocols, experimenters play a key role \cite{Sexton15}. For instance, they introduce the technology to the participants, provide the participants with advice regarding how they should perform the MI tasks and keep the participants motivated throughout the training. Nonetheless, social and emotional feedback were shown to have an impact on user experience, motivation and MI-BCI performances \cite{Pillette17}. Despite the main role that experimenters have in the experimental process and the literature regarding the impact of social and emotional feedback, no studies had yet been led in MI-BCI to evaluate the influence experimenters might have on their own experimental results.

Experimenter related biases are an important concern in other fields such as ethics and business \cite{miyazaki08}, social research \cite{rosnow97} or economic research \cite{zizzo10}. Literature from different fields states that the characteristics of the experimenter may consciously or unconsciously affect the responses, behavior and performance of the participants via direct and/or indirect interactions \cite{rosnow97}. For example, it has been shown that an “experimenter demand effect” occurs when participants unconsciously try to fit the appropriate image reflected by the experimenter’s behavior and therefore want to please and assist the experimenters in obtaining their expected results \cite{rosnow97}. 

Several studies investigating experimenters' influence suggested that gender-interaction could have an impact. For example, the interplay of participant's and experimenter's genders may shape the experimenter demand effect. When participants are instructed by an opposite-sex experimenter, they seem more likely to act in ways that confirm the experimenter’s hypothesis \cite{nichols08}. Also, men participants seem to elaborate more on autobiographical memory report with women experimenters than with men experimenters and more than women participants in general \cite{Grysman17}. Proxemics studies, which study the amount of space that people feel necessary to set between themselves and others, provide another example of gender interaction. Men participants seem to keep a shorter distance from women than from men \cite{uzzell06}. Interestingly, participants also prefer a larger comfort and reachability distance when facing a virtual man as compared to a virtual woman \cite{iachini16}. In a pain-related study, it was shown that men participants tend to report higher cold presser pain to a man experimenter than to a woman one \cite{levine91}. Studies revealed that there were no interactions in the physiological data between experimenter gender and participant gender, suggesting that men participants reporting lower pain report to women experimenters is probably due to psychosocial factors \cite{aslaksen07}. Another gender-related example would be that defensiveness is associated with greater relative left frontal activation in the presence of experimenters from the opposite-sex compared to experimenters from the same-sex \cite{kline02}. Thus, participants who work with an opposite-gender or same-gender experimenter can have different neurological responses, such as differences in their EEG recordings \cite{chapman18}. In a recent study examining the effect of psychosocial factors —particularly sex-related effects— on a given neurofeedback training (learning to modulate sensorimotor rhythm power and theta/beta power), women participants trained by women experimenters learned significantly less than their counterparts trained by men experimenters \cite{Wood18}.

These observations have led us to think that a gender-interaction could have an effect on MI-BCI experimental results. Yet, it has never been tested in BCI. The aim of our study was therefore to investigate if there was an influence of the experimenters’ gender depending on the participants’ gender on MI-BCI performances and progression (i.e., the evolution of performances). 

\mysection{Materials \& methods}

\mysubsection{Participants} 

Fifty-nine healthy MI-BCI naïve participants (29 women; age 19-59; $\bar{X}$=29; SD=9.318) completed the study. None of them reported a history of neurological or psychiatric disorder. Experimenters who conducted the study were six scientists (3 women; age 23-37; $\bar{X}$=29.2; SD=5.60) among which two experienced in BCI experimentation (1 woman) and four beginners who were trained to perform a BCI experiment beforehand. Each experimenter was randomly assigned to 10 participants (5 women and 5 men) they had never met before the session. 

Our study was conducted in accordance with the relevant guidelines for ethical research according to the Declaration of Helsinki. Both participants and experimenters gave informed consent before participating in the study. In order to avoid biased behavior, this study was conducted using a deception strategy, partially masking the purpose of the study. Participants were told that the study aimed at understanding which factors (unspecified) could influence BCI progress and/or performance. The study has been reviewed and approved by Inria’s ethics committee, the COERLE.

\mysubsection{Experimental protocol} 

Each participant participated in one MI-BCI session of 2 hours. The session was organized as follows: (1) consent form signature and completion of several questionnaires (around 20 min), (2) installation of the EEG cap (around 20 min), (3) six 7-minute runs during which participants had to learn to perform two MI-tasks, i.e., imagine right or left hand movements, (around 60 min, including breaks between the runs), (4) completion of post-session questionnaires (around 5 min) and (5) uninstallation and debriefing (around 10 min). 

During each run, participants had to perform 40 trials (20 per MI-task, presented in a random order), each trial lasting 8s. At t = 0s, an arrow was displayed on the screen. At t = 2s, an acoustic signal announced the appearance of a red arrow, which appeared one second later (at t = 3s) and remained displayed for 1.250s.The arrow pointed in the direction of the task to be performed, namely left or right to imagine a movement of the left hand or the right hand. Finally, at t = 4.250s, a visual feedback was provided in the shape of a blue bar, the length of which varied according to the classifier output. Only positive feedback was displayed, i.e., the feedback was provided only when the instruction matched the recognized task. The feedback lasted 3.75 s and was updated at 16Hz, using a 1s sliding window. After 8 seconds of testing, the screen turned black again. The participant could then rest for a few seconds, and a new cross was then displayed on the screen, marking the beginning of the next trial.

The training protocol used was the Graz protocol \cite{pfurtscheller01} which is divided into two steps: (1) training of the system and (2) training of the user. The first two runs were used as calibration in order to provide examples of EEG patterns associated with each of the MI tasks to the system. During the first two runs, as the classifier was not yet trained to recognize the mental tasks being performed by the user, it could not provide a consistent feedback. In order to limit biases with the other runs, e.g., EEG changes due to different visual processing between runs, the user was provided with an equivalent sham feedback, i.e., a blue bar randomly appearing and varying in length.

We respected the following recommendations: encourage the user to perform a kinesthetic imagination \cite{neuper05} and leave users free to choose their mental imagery strategy \cite{kober13}, e.g., imagining waving at someone or playing the piano. Participants were instructed to find a strategy for each task so that the system would display the longest possible feedback bar. Instructions were written in advance so that all the participants started with the same standardized information.

\mysubsection{Questionnaires} 

We assessed personality and cognitive profile for both experimenters and participants with the 5th edition of the 16 Personality Factors (16PF5) \cite{Cattell95}, a validated psychometric questionnaire to assess different aspects of personality and cognitive profile. This questionnaire identifies 16 primary factors of personality, such as anxiety or autonomy.
Participants also completed a mental rotation test measuring spatial abilities \cite{Vandenberg78}. 

\mysubsection{EEG Recordings \& Signal Processing} 

To record the EEG signals, 27 active scalp electrodes, referenced to the left ear, were used (Fz, FCz, Cz, CPz, Pz, C1, C3, C5, C2, C4, C6, F4, FC2, FC4, FC6, CP2, CP4, CP6, P4, F3, FC1, FC3, FC5, CP1, CP3, CP5, P3, 10-20 system).
Electromyographic (EMG) activity of the hands were recorded using two active electrodes situated 2.5cm below the skinfold on each wrists. Electrooculographic (EOG) activity of one eye was recorded using three active electrodes. Two, situated below and above the eye, aimed at recording vertical movements of the eye and one on the side aimed at recording horizontal movements. Physiological signals were measured using a g.USBAmp (g.tec, Austria), sampled at 256 Hz, and processed online using OpenviBE 2.1.0 \cite{Renard10}. 

To classify the two MI tasks from EEG data, we used participant-specific spectral and spatial filters. First, from the EEG signals recorded during the calibration runs, we identified a participant-specific discriminant frequency band using the heuristic algorithm proposed by Blankertz et al. in \cite{blankertz08} (Algorithm 1 in that paper). Roughly, this algorithm selects the frequency band whose power in the sensorimotor channels maximally correlates with the class labels. Here we used channels C3 \& C4 after spatial filtering with a Laplacian filter as sensorimotor channels, as recommended in \cite{blankertz08}. We selected a discriminant frequency band in the interval from 5 Hz to 35 Hz, with 0.5Hz large bins. Once this discriminant frequency band identified, we filtered EEG signals in that band using a butterworth filter of order 5.

Then, we used the Common Spatial Pattern (CSP) algorithm \cite{Ramoser00} to optimize 3 pairs of spatial filters, still using the data from the two calibration runs. Such spatially filtered EEG signals should thus have a band power which is maximally different between the two MI conditions. We then computed the band power of these spatially filtered signals by squaring the EEG signals, averaging them over a 1 second sliding window (with 1/16th second between consecutive windows), and log-transforming the results. This led to 6 different features per time window, which were used as input to a Linear Discriminant Analysis (LDA) classifier \cite{Lotte18}. As mentioned above, this LDA was calibrated on the data from the two calibration runs. These filters and classifier were then applied on the subsequent runs to provide online feedback. 

\mysubsection{Variables \& Factors} 

The aim was to evaluate the influence of the gender of the experimenters and participants on the MI-BCI performances of the participants over a series of 4 runs with online BCI use. Two measures were used to assess the performance of the participants. 

The first is the mean classification accuracy which is traditionally used by the community. This measure represents the percentage of time windows from the feedback periods that were correctly classified. However, this metric only considers whether the classification was correct, but not the quality of this classification, i.e., it does not take into account the classifier output. Since our participants were instructed to train to obtain not only a correct classification, but also a feedback bar as long as possible, we also studied a metric considering the feedback bar length, i.e., the classifier output.

Thus, we also used the Quality-Weighted Accuracy (QWA), the standard performance metric provided in OpenViBE MI-BCI applications, which is inspired by the SensoriMotor Rhythm quality score in \cite{GrosseWentrup12}. To compute it, we first summed the (signed) LDA classifier outputs (distance to the separating hyperplane) over all time windows during a trial feedback period. If this sum sign matched the required trial label, i.e. negative for left hand MI and positive for right hand MI, then the trial was considered as correctly classified, otherwise it was not. Finally, a run QWA was estimated as the percentage of trials considered as correctly classified using this approach.

\mysection{Results} 

\mysubsection{Comparability of groups}

Among 59 participants, 3 outperformed the others (by more than two SDs) both in term of mean classification accuracy (Outliers $\bar{X}$1=88.94, $\bar{X}$2=90.36, $\bar{X}$3=94.51; $\bar{X}$grp=59.33\%; SDgrp=12.3) and QWA (Respectively, outliers $\bar{X}$1=98.13, $\bar{X}$2=98.13, $\bar{X}$3=99.38; $\bar{X}$grp=62.78\%; SDgrp=16.2). Thus, the following analyzes are based on the results of 56 participants (27 women).

Before it all, we verified if the distribution of the data collected was normal using Shapiro-Wilk tests. The variables describing the mental rotation scores (p=0.34), anxiety (p=0.06) and autonomy (p=0.14) of our participants could be considered as having a normal distribution. Though, the mean classification accuracy of the runs did not have a normal distribution (p$\leq$\begin{math}10^{-3}\end{math}) and neither did the QWA metrics for the different runs (p$\leq$\begin{math}10^{-3}\end{math}).

We also checked that groups formed by participants' gender, i.e., \emph{``ParGender''}, and experimenters' gender, i.e., \emph{``ExpGender''}, were comparable. We focused on mental rotation scores (MRS), anxiety or autonomy, which were shown to influence on MI-BCI performances \cite{Jeunet15b}. Participants with low MRS \cite{Vandenberg78}, anxious or non-autonomous (both measured using the 16PF5 questionnaire \cite{Cattell95}) were shown to have lower MI-BCI performances than the others \cite{Jeunet15b, Tan14}. To check that groups were comparable, we ran 2-way ANOVAs with \emph{``ExpGender*ParGender''} as independent variables and either mental rotation scores, anxiety or autonomy as dependent variable. 

Results indicate that groups are comparable in terms of anxiety. Though, participants' gender influence their MRS [F(1,52)=17.47; p$\leq$\begin{math}10^{-3}\end{math}, $\eta^{2}$=0.25]. Men ($\bar{X}$men=0.072; SD=0.024) had higher MRS than women ($\bar{X}$women=0.045; SD=0.023), which is in accordance with the literature \cite{Linn85}. Furthermore, participants training with men or women experimenters did not have the same level of autonomy [F(1,52)=4.01; p=0.05, $\eta^{2}$=0.07]. Participants training with men experimenters ($\bar{X}$menExp=6.35; SD=1.74) were more autonomous than participants training with women experimenters ($\bar{X}$womenExp=5.67; SD=1.66). Therefore, we controlled for the influence of these variables in our subsequent analyses by using them as covariates in ANCOVAs (see paragraph \emph{Checking for confounding factors}).

\mysubsection{Participants' and experimenters' gender}

Then, we analyzed the influence of the gender of the experimenters and participants on the MI-BCI performances of the participants over the runs, i.e., \emph{``Run''}. To do so, we performed a 3-way repeated measures mixed ANOVAs with \emph{``ExpGender*ParGender*Run''} as independent variables and the repeated measures of performance over the runs, i.e., mean classification accuracy or QWA, as dependent variable. Even though the normality of the data is a pre-requisite of an ANOVA, the ANOVA is considered as robust against the normality assumption and to the best of our knowledge no other non parametric test enabled to perform such analysis. 

First, we performed such ANOVA using the mean classification accuracy. Results revealed no simple effect of \emph{``Run''} [F(3,156)=1.53; p=0.22, $\eta^{2}$=0.03], \emph{``ExpGender''} [F(1,52)=0.26; p=0.61, $\eta^{2}\leq$\begin{math}0.01\end{math}] and \emph{``ParGender''} [F(1,52)=0.23; p=0.64, $\eta^{2}\leq$\begin{math}0.01\end{math}]. They revealed no interaction of \emph{``Run*ParGender''} [F(3,156)=1.92; p=0.13, $\eta^{2}$=0.04], \emph{``Run*ExpGender''} [F(3,156)=0.23; p=0.87, $\eta^{2}$=\begin{math}0.01\end{math}] nor \emph{``ParGender*ExpGender''} [F(1,52)=0.92; p=0.34, $\eta^{2}$=0.02]. Finally, the interaction of \emph{``Run*ParGender*ExpGender''} was not significant either [F(3,156)=1.38; p=0.25, $\eta^{2}$=0.03].

\begin{figure}[h]
	\centering
	\includegraphics[width=8cm]{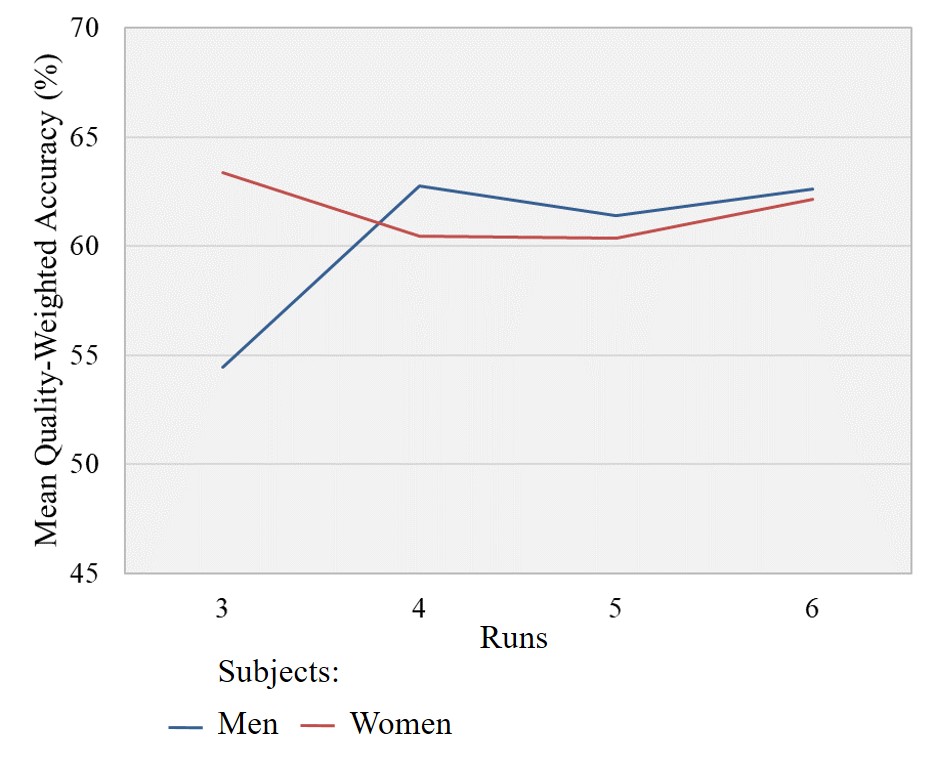}
	\caption{QWA evolution depending on participants' gender.} 
	\label{fig:QWA_subRun}
\end{figure}

Next, we performed this same analysis using QWA. Results revealed no simple effect of \emph{``Run''} [F(3,16)=1.81; p=0.15, $\eta^{2}$=0.03], \emph{``ExpGender''} [F(1,52)=0.54; p=0.47, $\eta^{2}$=0.01] nor \emph{``ParGender''} [F(1,52)=0.09; p=0.76, $\eta^{2}$=\begin{math}0.01\end{math}]. They also revealed no interaction of \emph{``Run*ExpGender''} [F(3,16)=0.08; p=0.97, $\eta^{2}$=\begin{math}10^{-2}\end{math}] nor \emph{``ParGender*ExpGender''} [F(1,52)=0.60; p=0.44, $\eta^{2}$=0.01]. Though, the \emph{``Run*ParGender''} interaction  was significant [F(3,156)=5.98; p=\begin{math}0.001\end{math}, $\eta^{2}$=0.1]. Figure \ref{fig:QWA_subRun} represents the evolution of the participants' QWA depending on their gender. 

\begin{figure}[h]
	\centering
	\includegraphics[width=8cm]{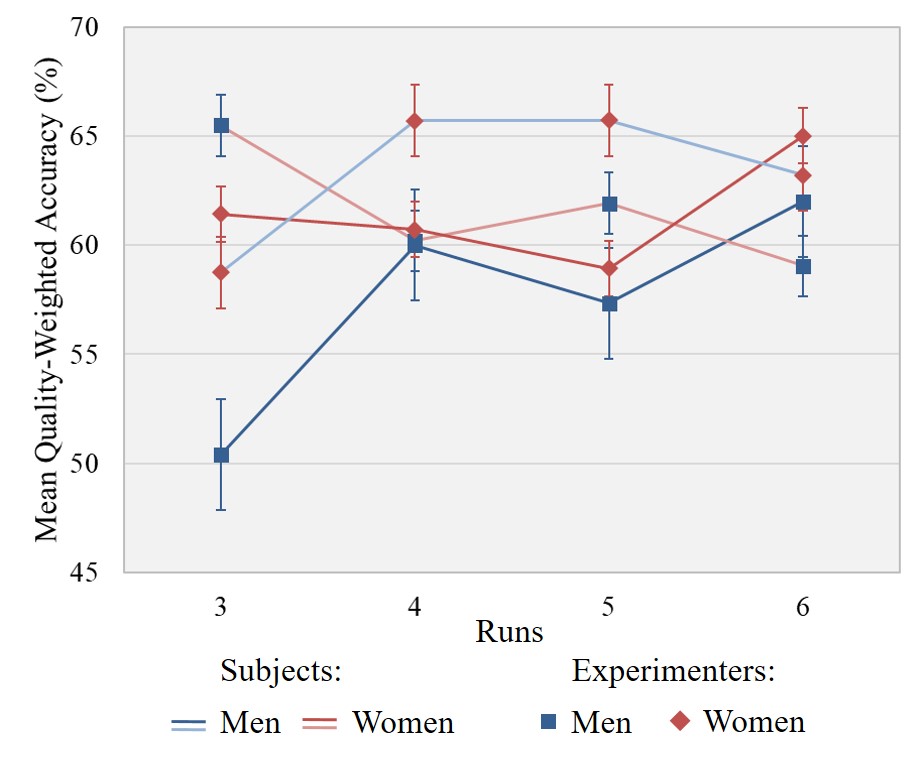}
	\caption{Participants' QWA evolution depending on the participants' and experimenters' gender.}
	\label{fig:QWA_expSubRun}
\end{figure}

Finally, a significant \emph{``Run*ParGender*ExpGender''} interaction was found [F(3,156)=3.46; p=0.02, $\eta^{2}$=0.06]. Figure \ref{fig:QWA_expSubRun} represents the participants' QWA evolution depending on the participants' and experimenters' gender. 

\mysubsection{Checking for confounding factors}

As stated before, the groups of participants formed using the participants' and experimenters' gender had differences in terms of mental rotation scores and autonomy. Therefore, we studied the potential impact these differences could have had on our results. First, we checked if a correlation could be found between our metrics of performances and these variables. No significant correlation was found between the autonomy and the mean classification accuracy ($r=-0.11$, $p=0.40$) nor the QWA ($r=-0.07$, $p=0.62$). The correlations between the mental rotation score and the mean classification accuracy ($r=-0.13$, $p=0.36$) or the QWA ($r=-0.24$, $p=0.08$) was not significant either.

Second, we ran a 3-way repeated measures mixed ANCOVA with \emph{``ExpGender, ParGender, Run''} as independent variables and one of the measures of performance, i.e., mean classification accuracy or QWA, as dependent variable, with the autonomy, i.e., \emph{``Aut''}, or the mental rotation score, i.e., \emph{``MRs''}, of the participants as covariate. 
When performing the analysis on the QWA we did not find any single effect or interaction of the autonomy (\emph{``Aut''} [F(1,51)=0.26; p=0.61, $\eta^{2}$=\begin{math}10^{-2}\end{math}], \emph{``Aut*Run''} [F(3,15)=0.81; p=0.49, $\eta^{2}$=0.02]) or the mental rotation score (\emph{``MRs''} [F(1,51)=1.75; p=0.19, $\eta^{2}$=0.03], \emph{``MRs*Run''} [F(3,15)=1.52; p=0.21, $\eta^{2}$=0.03]). 

When investigating the mean classification accuracy we found as well no impact of the autonomy (\emph{``Aut''} [F(1,51)=0.44; p=0.51, $\eta^{2}$=\begin{math}10^{-2}\end{math}], \emph{``Aut*Run''} [F(3,15)=1.46; p=0.23, $\eta^{2}$=0.03]) or the mental rotation score (\emph{``MRs''} [F(1,51)=1.05; p=0.31, $\eta^{2}$=0.02], \emph{``MRs*Run''} [F(3,15)=1.35; p=0.26, $\eta^{2}$=0.03]).

\mysection{Discussion} 

We analyzed results using two metrics of performances: QWA which represented what the participants were instructed to improve during training, and the mean classification accuracy, a traditional measure of BCI performances. Initial differences in mental rotation scores and autonomy between groups did not seem to bias results.

No influence of the experimenters’ and/or participants’ gender on the mean accuracy performance was found. Though, we found a significantly different evolution across runs of QWA between men and women participants (see Figure \ref{fig:QWA_subRun}). Women participants seemed to start the training with already good QWA, which decreased on the second run and increased again during the last run. Men participants however, started with rather low QWA and then drastically improved on the second run and then stagnated to reach slightly higher final QWA performances than women.

In addition, experimenters’ gender seemed to have an influence on this previous interaction. Indeed, the evolution of the QWA appear to depend on participants’ and experimenters’ gender (see Figure \ref{fig:QWA_expSubRun}). On the one hand, we found the same tendency for men participants to start with lower QWA at the beginning of the session independently of the experimenter’s gender. However, men seemed to start with drastically lower QWA performances when they were training with men experimenters. They also seemed to have higher QWA performances throughout the session when they were training with women experimenters. On the other hand, women participants seemed to start with higher QWA when training with men experimenters, though their QWA performances tended to drop throughout the session. However, when training with women experimenters they seemed to have a great increase in QWA during the last run.

Interestingly enough, this result does not match those of a recently published neurofeedback study \cite{Wood18}, in which the combination woman experimenter–woman participant appeared to hamper the training outcomes of the last, so that no learning effect was observed in this group.

\mysection{Conclusion} 

We investigated the potential influence of the experimenters’ gender depending on the participants’ gender on MI-BCI performances and progression throughout one MI-BCI session. Six experimenters (3 men; 3 women) trained 59 participants (30 men; 29 women). The general observation emerging from this study is that women experimenters seemed to induce better QWA performance progress for both men and women participants. Men participants seemed to start with substantially lower performances when they were training with men experimenters compared to when they were training with women experimenters. Also, even though women participants started with higher performances when training with men experimenters, their performances decreased throughout the session when they overall increased when training with women experimenters.

These results naturally need to be confirmed with larger populations. Further analysis are also needed regarding other variables that might influence or provide insights on our results. This includes inter-experimenter variables (e.g., traits or teaching competence), intra-experimenters variables (e.g., appearance or states), inter and intra-participants variables (e.g., traits or motivation) and interaction related variables (e.g., quantity and quality of interaction between the participant and the experimenter). There might also be other analysis to perform based on different performance metrics reflecting user performances independently of the classifier output \cite{Lotte18b}.

Further formal studies investigating the role of the BCI experimenter are needed. The need for research methods that explicit larger amounts of influencing factors (such as the experimenter) emerging from experimental protocol and context is equally important. In BCI research, the instructions (i.e., what participants are instructed to do during mental-imagery tasks) are rarely formalized, or in any case they are not taken into account and mentioned in papers. Similarly, protocols rarely evoke demonstrations \cite{Lotte13c} (i.e. showing a demonstration of a successful BCI use to the participant, together with a demonstration of feedback during (in)correctly performed mental tasks). 
It is common practice for studies in the BCI field not to report experimenter gender, though the literature as well as our results indicate that the influence of experimenters should be considered carefully while designing and reporting experimental protocols.

Literature suggests several solutions to limit the potential bias arising from the experimenter \cite{rosnow97, miyazaki08}. These methods include: monitoring participant-experimenter interaction; increasing the number and diversity of data collectors; pre-testing the method and controlling expectancy; providing an extensive training for administrators/ data collectors; monitoring and standardizing the behavior of experimenters with detailed protocol and pre-written instructions for the participant; and statistically controlling for bias. 

Beyond the potential bias that could arise from the experimenters' presence, the social and emotional feedback that experimenters provide could benefit MI-BCI. Indeed, the use of social feedback in BCI has been encouraged \cite{Sexton15}. Social presence and trust relationship between the user and the experimenter are essential for maintaining training motivation, which has been shown to facilitate the BCI learning process \cite{kleih11}. 

During MI-BCI training, using a learning companion to provide the participants with social and emotional feedback have proven effective in improving the user experience \cite{Pillette17}. An advanced conversational agent could also be used to supplement the role of the experimenter. It would represent yet another interesting method to control and/or enhance the experimenter influence.
Taking experimenter-related factors into account might lead to a conjoint progress of the global BCI performance and the validity and understanding of BCI experimental results.

\mysection{ACKNOWLEDGEMENTS}

This work was supported by the European Research Council (grant ERC-2016-STG-714567) and the French National Research Agency (grant ANR-15-CE23-0013-01). We express our gratitude to all the experimenters and participants who took part in this study.

\mysection{References}
\printbibliography[heading=none]
\end{document}